\documentclass{moriond}

\def\be{\begin{equation}}
\def\ee{\end{equation}}
\def\bea{\begin{eqnarray}}
\def\eea{\end{eqnarray}}

\def\babar{\mbox{\slshape B\kern-0.1em{\smaller A}\kern-0.1em
    B\kern-0.1em{\smaller A\kern-0.2em R}}}
\usepackage[export]{adjustbox}
\usepackage{relsize}
\usepackage{amsmath}
\newcommand{\Photo}{\includegraphics[height=35mm]}

\begin{document}
\vspace*{4cm}
\title{Recent Results from ~\babar: Dark Matter, Axion-like Particles and Heavy Neutral Leptons}

\author{ Sophie Charlotte Middleton\footnote{on behalf of the ~\babar~Collaboration} }

\maketitle

\abstract{Three independent searches for new physics using data collected at~\babar~are presented. Firstly, two searches for dark matter and baryogenesis: $B^{0}\rightarrow \Lambda + \psi_{D}$ and $B^{+} \rightarrow p + \psi_{D}$ are detailed, where $\psi_{D}$ is a new dark fermion. Neither signal is observed and new upper limits on the branching fractions, at the 90 $\%$ confidence level (C.L), are placed at  $\mathcal{O}(10^{-5} - 10^{-6})$ across the mass range $1.0< m_{\psi_{D}} <4.3$ GeV/c$^{2}$. Secondly, new limits on the coupling, $g_{aW}$, of an axion-like particle ($a$) to the $W$ boson, at the 90 $\%$ C.L, are presented at $\mathcal{O}(10^{-5})$ GeV$^{-1}$ for $a$ masses in the mass range 0.175 $<m_{a}<$ 4.78 GeV/c$^{2}$. Thirdly, a model-independent search for heavy neutral leptons (HNL) found new upper limits at the 95 $\%$ C.L on the extended Pontecorvo–Maki–Nakagawa–Sakata (PMNS) matrix element, $|U_{\tau 4}|^{2}$, which depend on the HNL mass hypothesis and vary from $2.31 \times 10^{-2}$ to  $5.04 \times 10^{-6}$, across the mass range $100 < m_{4} < 1300 ~\text{MeV}/c^{2}$, with more stringent limits on higher HNL masses.}

\section{Motivations: the need for new physics}

Since its formulation, the Standard Model (SM) has successfully explained many experimental results. The particles and the 18 free parameters have been tested with extraordinary precision, in some cases, up to 1 part in a
trillion\cite{PhysRevLett.100.120801}. However, there remains plenty of data that the SM cannot fully
explain. These include: the origins of the neutrino masses and neutrino mixing; the existence of large amounts of non-baryonic cold dark matter in the Universe; and, the observed dominance of matter over anti-matter.

There are also theoretical motivations to search for physics beyond the SM (BSM), including:
to understand the strong CP problem; to try to unify gravity with the other fundamental forces; to account for Higgs mass hierarchy problem; and, to understand the structure of fermion masses and mixings and the number of fermion families.


In this article, three independent searches for new physics which can explain several of the above-described phenomena are detailed. All three analyses utilize data collected at ~\babar.

\section{The BaBar Detector}

The data sample used in these analyses corresponds to a total integrated luminosity of 431 fb$^{-1}$, collected with the~\babar~ detector at the PEP-II $e^{+}e^{-}$ storage ring at the SLAC National Accelerator Laboratory. At PEP-II, 9.0 GeV electrons collide with 3.1 GeV positrons at center-of-mass (CM) energies near 10.58 GeV, on the $\Upsilon(4S)$ resonance.

\begin{figure}[h]
\centering
        \includegraphics[width=3.0in]{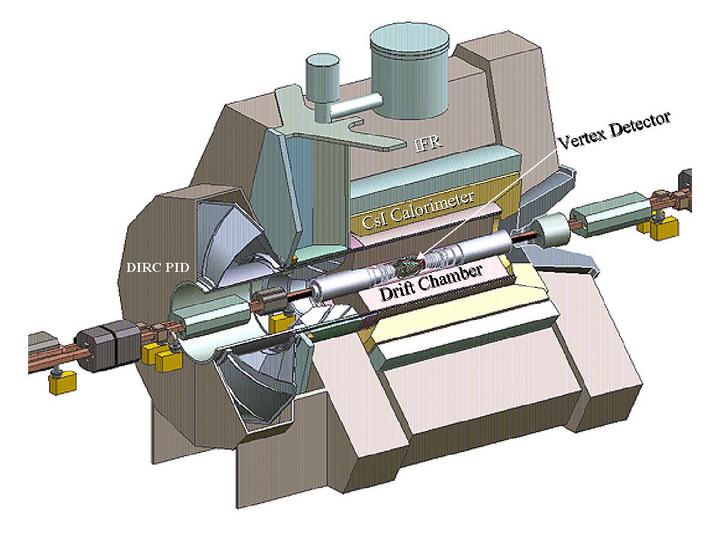}
        \caption{The~\babar~detector system.}
        \label{fig:babar}
\end{figure}

In the \babar~ detector\cite{BaBar} (see Fig.~\ref{fig:babar}), a silicon vertex tracker (SVT) and a 40 layer Drift Chamber (DCH), placed inside a 1.5-T solenoid magnet, are utilized to reconstruct charged-particle tracks. The transverse momentum resolution is $0.47\%$ at 1 GeV/$c$, where the transverse momentum, $p_{T}$, is defined as the total momentum of all four tracks orthogonal to the beam axis. An Electro-Magnetic Calorimeter (EMC) measures the energy of electrons and photons with a resolution of $3\%$ at 1 GeV. A Ring Imaging Cherenkov detector (DIRC) is located in front of the EMC and is used, together with specific ionization loss measurements in the SVT and DCH, to identify charged  pions and kaons, and provide additional electron identification. The instrumented flux return of the solenoid is used to identify muons.

\section{Search for Dark Matter and Baryogenesis}

The existence of dark matter  is long established from cosmological observations. Precision measurements of the Cosmic Microwave Background (CMB) by the Planck satellite\cite{refId0} suggest only $\sim$15$\%$ of the Universe's matter is SM particles. The remaining fraction is referred to as dark matter (DM). Establishing the mass-scale and nature of DM is one of the main goals of modern experimental particle physics. Another is understanding the baryon asymmetry of the Universe (BAU)\cite{Canetti_2010}. Baryogenesis is needed to produce an initial excess of baryons over anti-baryons consistent with both CMB and big-bang nucleosynthesis (BBN) measurements\cite{PDG2020,RevModPhys.88.015004}.

Elor \textit{et al.}\cite{Elor:2018twp} propose the existence of a new dark sector anti-baryon, $\psi_{D}$, which is a DM candidate. The proposed model also explains the BAU.  Baryogenesis occurs due to out-of-thermal equilibrium production of $b$ and $\bar{b}$ quarks in the early Universe. These quarks then hadronize to form  $B^{0}$ and $B^{\pm}$ mesons. The $B^{0} - \bar{B}^{0}$ mesons then undergo CP-violating oscillations before decaying into a SM baryon, $\psi_{D}$, and possibly any number of additional light SM mesons. Matter-antimatter asymmetries result in both the visible and dark sectors with equal magnitudes but opposite signs. Thus total baryon number is conserved. 

\begin{figure}[t]
    \centering
    \includegraphics[width=0.415\textwidth]{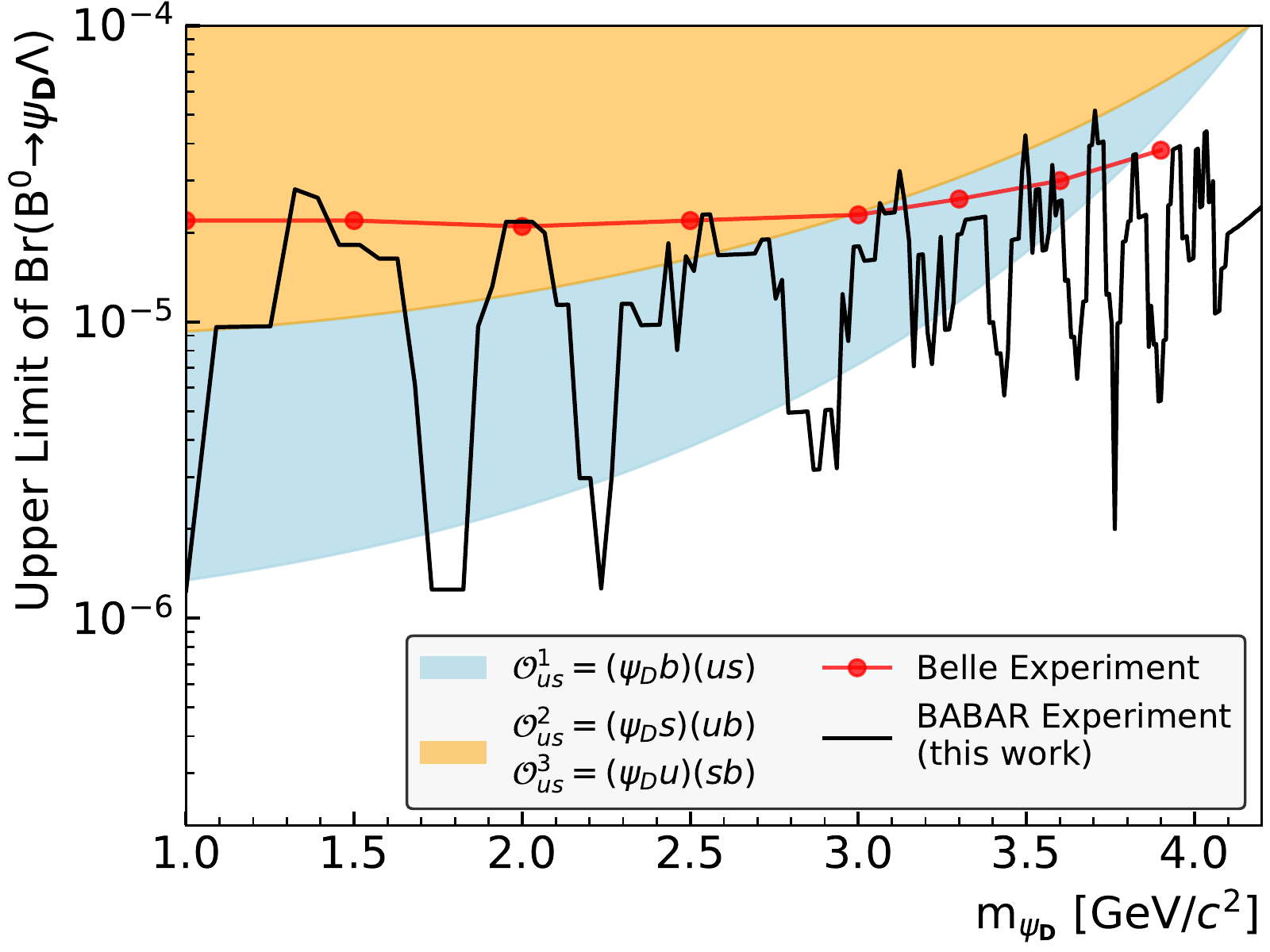}
    \includegraphics[width=0.42\textwidth]{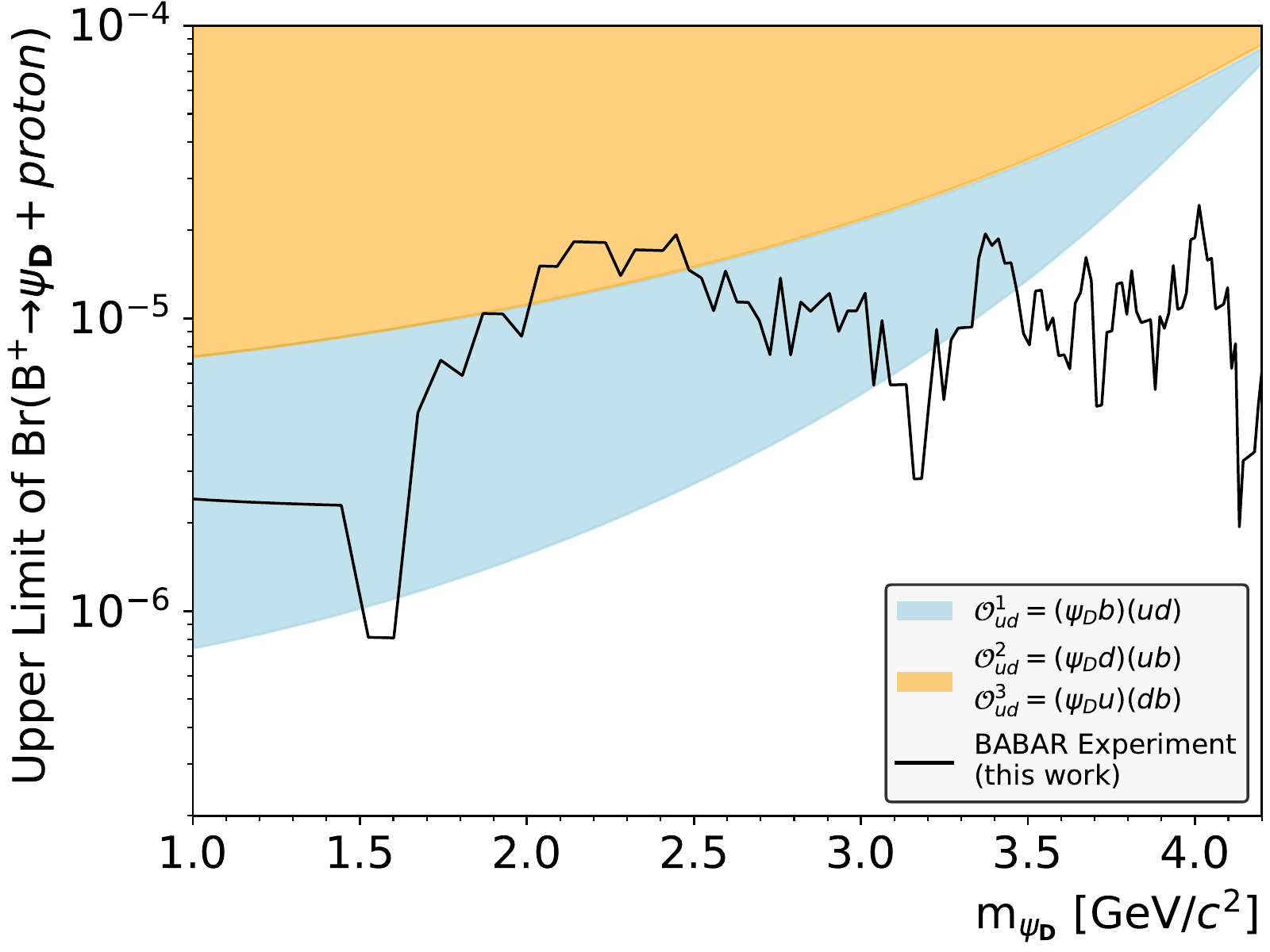}
    \caption{Derived 90 $\%$ C.L upper limits on the branching fraction (left) $B^{0} \rightarrow \psi_{D} + \Lambda$  (right) $B^{+} \rightarrow \psi_{D} + p$  for ~\babar~ data set corresponding to 398 fb$^{-1}$. The theory expectation values for the effective operators were received from Elor \textit{et al.} following communication with the authors.}
    \label{fig:DMlimit}
\end{figure}

$B$-factories are  an ideal place to search for $\psi_{D}$ through missing mass associated with $B$ meson decays. ~\babar~ recently published a search for $B^{0} \rightarrow \Lambda + \psi_{D}$\cite{BaBar:2023rer}, Fig.~\ref{fig:DMlimit} (left) shows the resulting 90$\%$ confidence level (C.L) upper limit on the branching fraction along with theory predictions. Figure~\ref{fig:DMlimit} (right) shows an analogous plot for the $B^{+} \rightarrow p + \psi_{D}$ channel. Tight constraints are provided, ruling out a large amount of parameter space in both cases, and almost fully excluding some operators. In both analyses, a scan method was utilized in which a set of 8 Monte Carlo (MC) signal samples across the mass range 1.0$< m_{\psi_{D}}<$ 4.3 GeV/c$^{2}$ were made using EvtGen\cite{EvtGen}. The tagging hadronic recoil method was utilized in which one of the outgoing B mesons, termed the $B_{tag}$, is fully reconstructed and the second is termed the $B_{sig}$. The missing mass ($m_{miss.}$), which in the case of a signal would be the $\psi_{D}$ mass, is calculated from the four momenta of $B_{sig}$ and baryon $\mathcal{B}$:

\begin{equation}
    m_{miss.} = \sqrt{(E^{\:*}_{B_{sig}} - E^{\:*}_{\mathcal{B}})^{2} - | \vec{p}^{\:*}_{B_{sig}} - \vec{p}^{\:*}_{\mathcal{B}} |^{2} }
\end{equation}

where ($\vec{p}^{\:*}_{B_{sig}},E^{\:*}_{B_{sig}}$) and ($\vec{p}^{\:*}_{\mathcal{B}},E^{\:*}_{\mathcal{B}}$) are the four-momenta of the signal $B_{sig}$ and SM baryon in the CM frame, respectively.  A scan is performed across the missing mass distribution with a step size equal to the signal mass resolution ($\sigma_{m}$) interpolated from fits to the signal MC samples. In total $\mathcal{O}(100)$ mass hypotheses were considered in both cases across the range 1.0 $<m_{miss.} <$ 4.3 GeV/c$^{2}$ for both channels.

The signal MC samples were used to help define a set of signal selection criteria, which were later used to define signal and background (side-band) regions in the missing mass distribution from the data. In the absence of a signal, the 90$\%$ C.L upper limits on the branching fractions are derived using a profile likelihood method\cite{Rolke_2005}. A Poisson counting approach is used where the number of signal and background events is assumed to follow Poisson distributions,
and the efficiency is assumed Gaussian with variance equal to the total systematic uncertainty. For a given $\psi_{D}$ mass hypothesis, the signal region is defined as the region $m_{\psi_{D}} - 5\sigma_{m}< m_{miss.} <  m_{\psi_{D}} + 5\sigma_{m}$, the side-bands ($[+5\sigma,+10\sigma]$ and $[-10\sigma,-5\sigma]$) on either side of this window are classified as the background region. Events in the data in each region are counted and input into the profile likelihood fit.

\section{Search for Axion-like Particles}

Many BSM theories include anomalous global symmetries whose spontaneous breaking leads to  axion-like particles (ALPs), and new pseudo-Goldstone bosons\cite{Peccei:1977hh,Weinberg:1977ma}. ALPs can be cold dark matter candidates\cite{Nomura:2008ru}. ALPs could also resolve several outstanding issues related to naturalness .e.g. the strong CP problem or the hierarchy problem. 

A ubiquitous feature of models predicting ALPs is their proposed coupling to pairs of SM gauge bosons. The coupling to $W$ bosons has much weaker experimental constraints than that to photons or gluons. Coupling to $W$ bosons would lead to ALP production in favor-changing neutral-current decays, such as B meson decays. The~\babar~ search\cite{BaBar:2021ich} presents the first search  ALPs produced in B meson decays and visibly decaying to two photons.

A minimal ALP ($a$) model is considered,  with coupling $g_{aW}$ to the $SU(2)_{W}$ gauge-boson field
strengths, $W^{b}_{\mu \nu}$, and Lagrangian:

\begin{equation}
    \mathcal{L} = -\frac{g_{aW}}{4}a W^{b}_{\mu \nu} \Tilde{W}^{\mu \nu}_{b}
\end{equation}

where $\Tilde{W}^{\mu \nu}_{b}$ is the dual field-strength tensor. This leads to the production of ALPs at one loop in the channel $B^{\pm} \rightarrow K^{\pm}a$, with the ALP being emitted from an
internal $W$ boson. The branching fraction of the di-photon decay  is nearly 100$\%$ for $m_{a} < m_{W}$. 

A search for $B^{\pm} \rightarrow K^{\pm}a$ followed by $a\rightarrow \gamma \gamma$ is conducted in the range 0.175 GeV/c$^{2}$ $< m_{a} < m_{B^{+}} - m_{K^{+}} \sim $ 4.78 GeV/c$^{2}$, excluding the mass intervals 0.45-0.63 GeV/c$^{2}$ and 0.91-1.01 GeV/c$^{2}$ due to backgrounds from $\eta$ and $\eta^{'}$ mesons, respectively. 

A scan method is employed, this time across the di-photon reconstructed invariant mass distribution, $m_{\gamma \gamma} $. Signal yields of promptly decaying ALPs are extracted  by performing a series of unbinned maximum likelihood fits of a hypothetical signal peak over a smooth background. In total 461 signal mass hypotheses are considered with scan step size determined from the signal resolution, $\sigma_{\gamma \gamma}$. 

In the absence of a significant signal, Bayesian upper limits at 90$\%$ C.L on  $B^{\pm} \rightarrow K^{\pm}a$ and $a\rightarrow \gamma \gamma$ are derived with a uniform positive prior in the product branching fraction systematic effects are included by convolving likelihood with a Gaussian distribution whose standard deviation is equal to the systematic uncertainty. 

In addition to the prompt process ALPs can be long-lived at small masses and coupling,  ALP proper decay lengths of $c\tau_{a}$ = 1, 10, and 100 mm are considered for masses $m_{a} < 2.5$ GeV$^{2}$. The $c\tau_{a}$ dependence of the limit is found to be less pronounced at higher masses because the ALP is less boosted, having a shorter decay length in the detector. An interpolating function is developed to obtain product branching fraction limits for intermediate lifetimes.

\begin{figure}[t]
    \centering
    \includegraphics[width=3.5in]{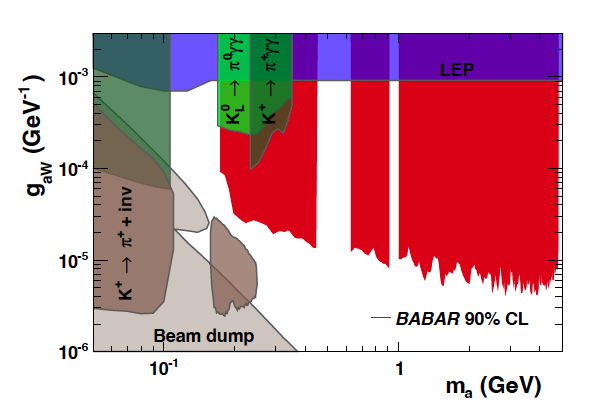}
    \caption{Derived 90 $\%$ C.L. upper limits on the ALP-$W$ boson coupling, $g_{aW}$. ~\babar~ data corresponds to 424 fb$^{-1}$. }
    \label{fig:alplimit}
\end{figure}

 We can use the branching fraction as a function of the lifetime to set limits on the coupling. Figure~\ref{fig:alplimit} shows the 90$\%$ C.L bounds on $g_{aW}$ which are $\mathcal{O}(10^{-5})$ GeV$^{-1}$ across the mass range, improving on previous constraints by more than two orders of magnitude.

\section{Search for Heavy Neutral Leptons}


Heavy Neutral Leptons (HNLs) have long been used to explain the smallness of neutrino mass\cite{small} and provide a dark matter candidate\cite{DM1,DM2}. HNLs could also be responsible for the generation of the matter-antimatter asymmetry of the Universe via leptogenesis\cite{baryo}. The Neutrino Minimal Standard Model ($\nu$-MSM) \cite{neut_min} proposes three additional heavy neutrinos. Two of the additional neutrinos have masses in the MeV/$c^{2}$ - GeV/$c^{2}$ range; the third is a keV/$c^{2}$ scale dark matter candidate.

The~\babar~analysis \cite{BaBar:2022cqj} strategy relies on the idea that an HNL can interact with the tau via a charged-current weak interaction. If the visible decay products of the $\tau$ have recoiled against a heavy neutrino, the kinematics of these particles are modified with respect to SM $\tau$ decay with a massless neutrino. It is assumed that the HNL does not decay within the detector.

The~\babar~search studies only the 3-prong charged pion $\tau$ decay, giving access to the region 300$<m_{4}<$1360 MeV$/c^{2}$, which historically has weaker constraints. Combining the three charged pions into a hadronic system, the decay can be considered two-bodied, with a heavy neutrino recoiling against a hadornic system. The technique used to place upper limits on $|U_{\tau 4}|^{2}$ proceeded using Monte Carlo simulation to construct 2D template histograms of the invariant mass and energy of the visible system.

\begin{figure}[h]
         \centering
         \includegraphics[width=3.5in]{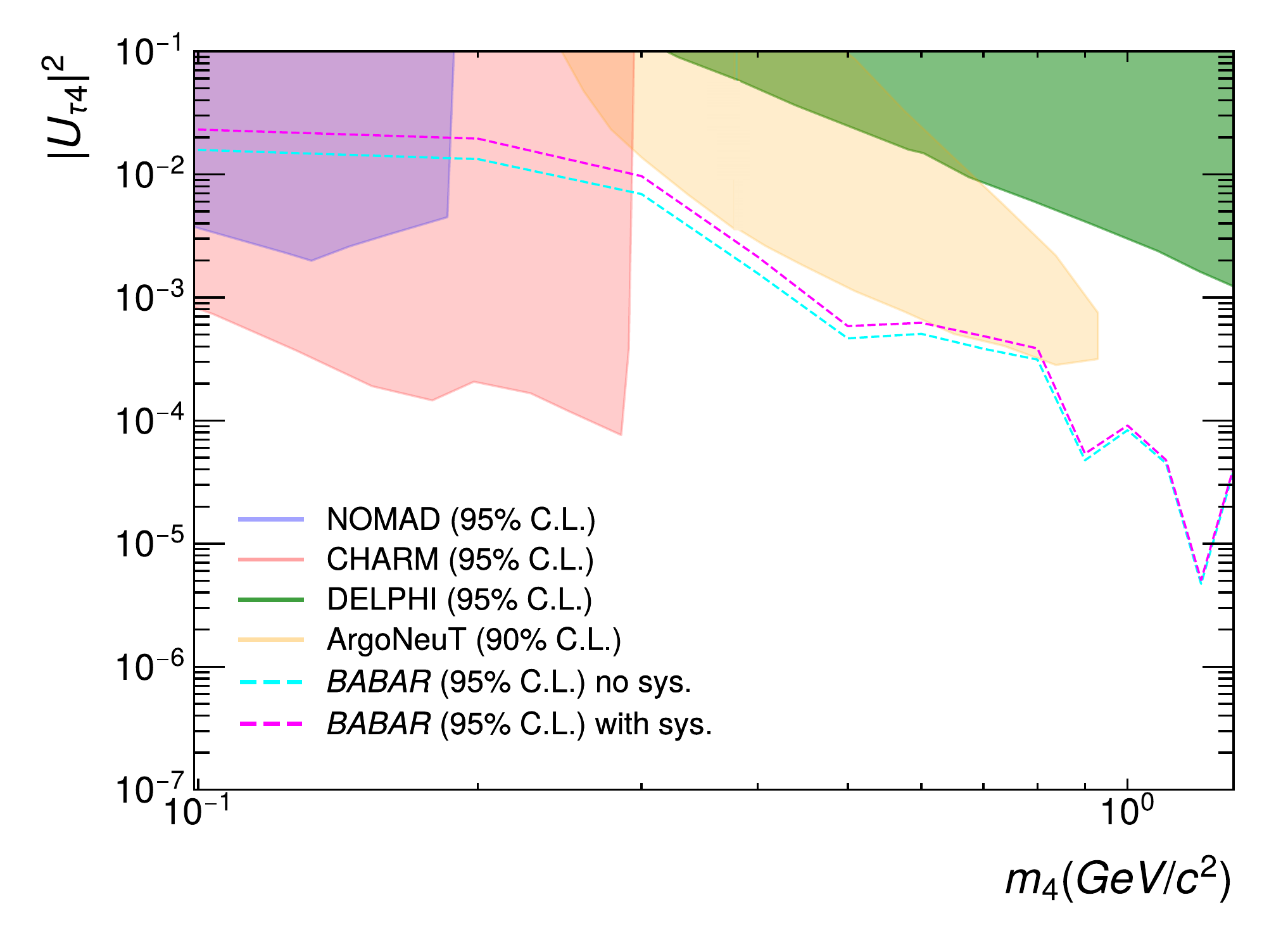}
        \caption{Upper limits at 95$\%$ C.L on $|U_{\tau 4}|^{2}$. The magenta line represents the result when uncertainties are included this is expected to be a very conservative upper limit. ~\babar~ data corresponds to 424 fb$^{-1}$. }
        \label{fig:limits}
\end{figure}

Figure~\ref{fig:limits} shows the results from the presented analysis. A model-independent search for heavy neutral leptons (HNL) found new upper limits at the 95 $\%$ C.L on the  $|U_{\tau 4}|^{2}$ which vary from $2.31 \times 10^{-2}$ to  $5.04 \times 10^{-6}$, across the mass range $100 < m_{4} < 1300 ~\text{MeV}/c^{2}$. More stringent limits are placed on higher neutrino masses. These improve on the previous upper limits from NOMAD \cite{Nomad}, CHARM \cite{CHARM} and DELPHI \cite{DELPHI}. In 2021 the ArgoNeuT experiment\cite{PhysRevLett.127.121801} also published limits in this region, the ~\babar~ result improves on those limits. In a recent publication Barouki \textit{et al.} \cite{10.21468/SciPostPhys.13.5.118} showed even tighter bounds. In addition, a recent recasting of the CHARM data, which utilizes data from the electron and muon searches to indirectly constrain the tau sector, has also improved constraints in the same region \cite{Boiarska_2021}. Constraints also exist from cosmic surveys for eV-scale Seesaw \cite{Canetti_2010}  and Big-Bang Nucleo-synthesis (BBN) \cite{BBN}. 

\section{Conclusions}

Three independent searches for new physics which can explain several beyond SM phenomena  are presented. All three show improved upper limits on the existence of new, beyond Standard Model, particles. In all cases, no new physics is observed but the new limits help constrain theoretical models.

\section*{Acknowledgments}

We are grateful for the extraordinary contributions of our PEP-II colleagues in achieving the excellent luminosity and machine conditions that have made this work possible. The success of this project also relies critically on the expertise and dedication of the computing organizations that support ~\babar. The collaborating institutions wish to thank SLAC for its support and the kind hospitality extended to them. We also wish to acknowledge the important contributions of J.~Dorfan, W.~Dunwoodie, and our deceased colleagues E.~Gabathuler, W.~Innes, D.W.G.S.~Leith, A.~Onuchin, G.~Piredda, and R. F.~Schwitters.

\section*{Appendix}

\bibliography{references}

\end{document}